\documentclass{llncs}
\usepackage{etex}

\usepackage{proof}
\usepackage{mathptmx}
\usepackage{hyperref}
\usepackage[utf8]{inputenc}
\usepackage{amsmath}
\usepackage{amssymb}
\usepackage{graphicx}
\usepackage{booktabs}
\usepackage[colorinlistoftodos]{todonotes}
\usepackage{xspace}
\usepackage{paralist, tabularx}
\usepackage{listings}
\usepackage{subcaption}
\usepackage{enumitem}

\def\hb{\hbox to 10.7 cm{}}

\newcommand{\Paragraph}[1]{\smallskip\noindent{\bf #1.}}

\newcommand{\sysstat}{\mathit{Sys}}
\newcommand{\closecomp}{\mathit{close(C)}}
\newcommand{\opencomp}{\mathit{open(C)}}
\newcommand{\tinc}{\mathit{heat_1(C)}}
\newcommand{\pinc}{\mathit{heat_2(C)}}
\newcommand{\ptinc}{\mathit{heat_3(C)}}
\newcommand{\damage}{\mathit{damaged(C)}}
\newcommand{\dya}{{\mathit DY}}

\newcommand{\inv}[1]{\mathit{inv}(#1)}
\newcommand{\scrypt}[2]{\{\!\mid\!\! #1 \!\!\mid\!\}_{#2}}
\newcommand{\acrypt}[2]{\{#1\}_{#2}}
\newcommand{\conc}[1]{[#1]}
\newcommand{\dy}{DY}
\newcommand{\DYComp}{G}
\newcommand{\DYCompAxiom}{\DYComp_{\mathrm{axiom}}}
\newcommand{\DYCompPair}{\DYComp_{\mathrm{pair}}}
\newcommand{\DYCompCrypt}{\DYComp_{\mathrm{crypt}}}
\newcommand{\DYCompScrypt}{\DYComp_{\mathrm{scrypt}}}

\newcommand{\DYPK}{A_{\mathrm{crypt}}}
\newcommand{\DYPrK}{A^{-1}_{\mathrm{crypt}}}
\newcommand{\DYSK}{A_{\mathrm{scrypt}}}

\newcommand{\DYCi}{A_{\mathrm{pair}_i}}

\lstdefinelanguage{aslanpp}{
  basicstyle=\normalsize\ttfamily,
  breakatwhitespace=false,
  xleftmargin=10pt,
  breaklines=true,
  mathescape=true,
   numbers=left,
   numberstyle=\tiny,
  morekeywords={specification,channel_model,CCM,ICM,ACM,entity,on,else,import,types,symbols,nonpublic,noninvertible,macros,clauses,equations,body,breakpoints,new,any,where,send,receive,over,retract,if,then,elseif,while,select,on,assert,constraints,goals,forall,exists,Actor,for} 
}
\begin{document}

\title{CPDY: Extending the Dolev-Yao Attacker with Physical-Layer Interactions}

\author{
Marco Rocchetto\inst{1}\thanks{The work was carried out while Marco was with iTrust at Singapore University of Technology and Design} \and Nils Ole Tippenhauer\inst{2}
}

\institute{
SnT, University of Luxembourg \and
ISTD, Singapore University of Technology and Design
}

\maketitle


\begin{abstract}
  We propose extensions to the Dolev-Yao attacker model to
  make it suitable for arguments about security of Cyber-Physical
  Systems. The Dolev-Yao attacker model uses a set of rules to define
  potential actions by an attacker with respect to messages
  (i.e. information) exchanged between parties during a protocol
  execution. As the traditional Dolev-Yao
  model considers only information (exchanged over a channel
  controlled by the attacker), the model cannot directly be used to
  argue about the security of cyber-physical systems where
  physical-layer interactions are possible.
  Our Dolev-Yao extension, called cyber-physical Dolev-Yao (CPDY) attacker
  model, 
  allows additional orthogonal interaction channels between the parties. In
  particular, such orthogonal channels can be used to model
  physical-layer mechanical, chemical, or electrical interactions
  between components. In addition, we discuss the inclusion of
  physical properties such as location or distance in the rule set. We
  present an example set of additional rules for the Dolev-Yao
  attacker, using those we are able to formally discover physical
  attacks that previously could only be found by empirical methods or
  detailed physical process models.
\end{abstract}


\section{Introduction}
\label{sec:introduction}
In recent years, security of Cyber-Physical systems (CPS) has received
increasing attention by researchers from the domain of computer
science, electrical engineering, and control
theory~\cite{mo12smartgrid,spacios3.3.2deliverable}.
We use the term CPS to refer to systems that consist of networked
embedded systems, which are used to sense, actuate, and control
physical processes. Examples for such CPS include industrial water
treatment facilities, electrical power plants, public transportation
infrastructure, or even smart cars. All those systems have seen a
rapid increase in automation and connectivity, which threatens to
increase vulnerability to malicious attacks.

Security analysis of any system relies on well-defined attacker and
system models~\cite{dolev83attacker,armando12avantssar}. 
While the system model
provides an appropriate abstraction of the system under attack, the
attacker model ideally fully defines the possible interactions between
the attacker and the attacked system. In particular, the model will
also define constraints for the attacker (e.g. finite computational
resources, no access to shared keys).

In contrast to the domain of information security, where the Dolev-Yao
attacker model~\cite{dolev83attacker} (DY) is widely used for
protocol analysis, the state-of-the-art for CPS security does not have
a common terminology for attacker models.  Even if the topic has been
broadly discussed in the CPS research community, e.g.,
in
~\cite{cardenas09challenges},
only a small number of tentative
works (e.g.,~\cite{vigo12attacker,leMay11advise}) have addressed that
problem. The DY model used by the information security community
represents a very strong attacker, who can access and manipulate all
network traffic arbitrarily. One could directly translate this
attacker to CPS by allowing the attacker to intercept any
communication in a real system (e.g. local fieldbus communication), or
to be within physical proximity of all (unprotected) devices.
However, such an attacker would only be capable of finding attacks on
the network level of the CPS. Since the network traffic of CPS
does not contain information about all interactions possible in a CPS,
it is not sufficient for comprehensive analysis. As a result,
there likely are (physical-layer) interactions between the attacker and the
system that cannot be captured by the DY paradigm.
 


In this paper, we investigate the application of the DY 
attacker model for security analysis of CPS. We present a set of
extensions to allow for a more general attacker model for CPS, that we
named CPDY (Cyber-Physical Dolev-Yao)~\cite{rocchetto16cpdytool}.

We summarize our contributions as follows:
\begin{compactitem}
\item We discuss the general limitations of the DY attacker model for analysis of CPS, and physical layer interactions between the attacker and the attacked system.
\item We propose a number of rule extensions to analyze CPS using the DY model.
\item We implement these rule extensions in the ASLan++~\cite{ASLan++-FMCO10} formal language, and present use case examples.
\end{compactitem}

\Paragraph{Structure} In Section~\ref{sec:background}, we summarize
the DY attacker model
.  We discuss the use of the DY model in the context of
CPS in Section~\ref{sec:motivation}, and show that the traditional
attacker and system model is only able to represent a subset of
possible interactions.  We propose extensions to the DY 
attacker model in Section~\ref{sec:extension}, and show 
our results on a real word water treatment plant use case in Section~\ref{sec:usecase}. 
We summarize the related work in Section~\ref{sec:related} and we conclude the paper in
Section~\ref{sec:conclusion}.

\section{Background}
\label{sec:background}
\subsection{Modeling Systems and Communications}
\Paragraph{Level of Modeling Detail} Formal languages,
e.g.  HLPSL~\cite{von2005high} and
ASLan++~\cite{ASLan++-FMCO10}, permit a modeler to define not only the exchanged messages but
also the behavior of entities involved in the
communication. 
Some of the security validation tools allow a modeler to benefit of
some algebraic properties
(e.g.,~\cite{schmidt14tamarin,BasinMoedersheimViganoIJIS05}) but these
are typically represented symbolically by a set of constraints.
Intuitively, a high level of details (e.g., a concrete highly detailed
representation of the behavior of an agent) may result in
non-termination problems while performing the analysis. In addition,
even if we could afford such a level of details, it might not be
useful to analyze security protocols against security properties at
that level of detail.  Some of the most common attacks (such as
man-in-the-middle and replay attacks) that violate confidentiality or
authentication can be found without the need of detailing the
encryption scheme in the protocol.

\Paragraph{Modeling Simplifications} In the so called \emph{perfect
cryptography} assumption, the security encryption scheme is suppose to
be ``perfect'', without any exploitable flaw, and so the only way for
the attacker to decrypt a message is by using the proper key.  That
assumption is widely accepted in the security protocol community, and
most of the formal reasoning tools for the analysis of security
protocols abstract away the mathematical and implementation details of
the encryption scheme~\cite{turuani06atse,BasinMoedersheimViganoIJIS05,ArmandoCompagna-jelia04-sd,rocchetto13spim}.

\Paragraph{Modeling Architectures} Following the same line of
reasoning, when one considers more complicated architecture
representation, such as web applications or Service Oriented
Architecture (SOA), some of the components are commonly assumed to be
``perfect'', in the same way as encryption schemes. Several examples are
shown by the case studies of the AVANTSSAR and SPaCIoS
projects~\cite{armando12avantssar,vigano13spacios} where researchers have
developed several abstract models of SOAs and web applications using
the ASLan++
language, e.g,  
in~\cite{avantssar-deliverable-5.3}. 
In order to give some specific examples, databases in SQLi analysis
in~\cite{buchler14inference}, CSRF token generation in~\cite{csrf} are
assumed to be ``perfect''.  This is due to the fact that (as the
security of security protocols is not guaranteed only by encryption
schemes) the security of web applications is not guaranteed only by
store procedures or perfect random generation of token.  Another
example has been presented in~\cite{camenisch10idmx}, where authors
search for attacks in zero-knowledge proof systems abstracting away
some of the mathematical and implementation details of the
zero-knowledge algorithms.


\subsection{Cyber-Physical Systems}
\begin{figure}[t]
\centering
\begin{subfigure}[t]{0.5\linewidth}
\includegraphics[width=\linewidth]{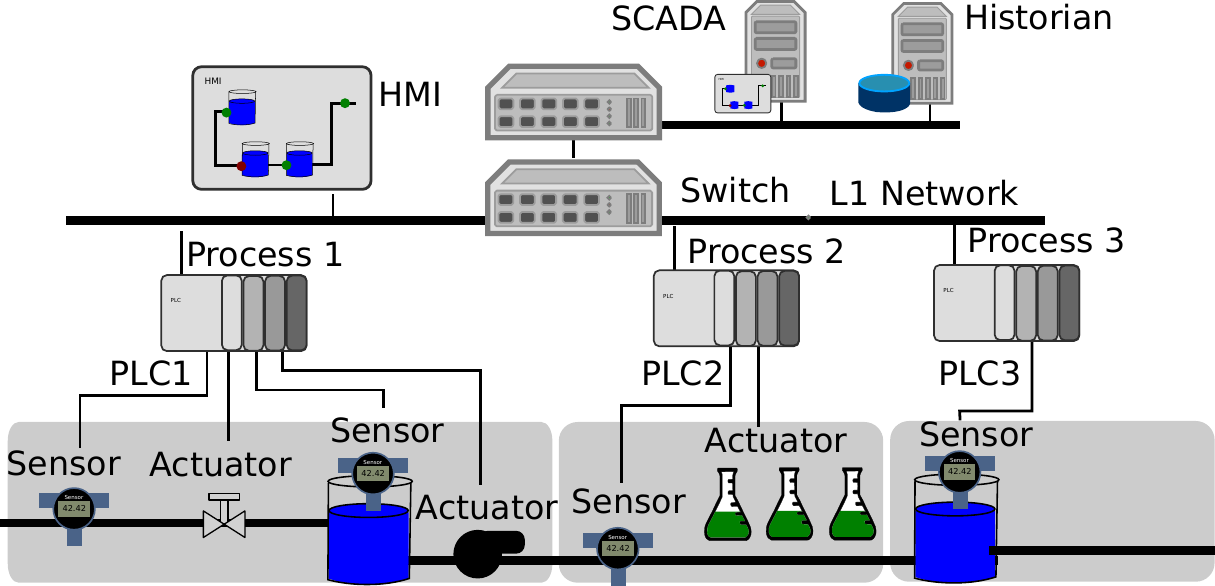}
\caption{}
\label{fig:swatNetwork}
\end{subfigure}
~
\begin{subfigure}[t]{0.4\linewidth}
\centering
\includegraphics[width=0.9\linewidth]{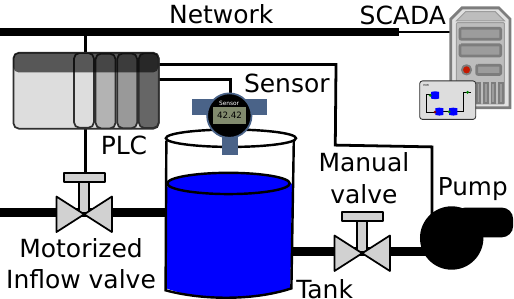}
\caption{}
 \label{fig:process}
\end{subfigure}
\caption{(a) Example CPS architecture. (b) Use case scenario of water tank with motorized valve and pump controlled by a PLC. A level meter reports to the PLC. A manual valve is placed between tank and pump.}
\end{figure}

In this work, we use the term Cyber-Physical System (CPS) to refer to
systems that consist of networked embedded systems, which are used to
sense, actuate, and control physical processes. Examples for such CPS
include industrial water treatment facilities, electrical power
plants, public transportation infrastructure, or even smart cars. All
those systems have seen a rapid increase in automation and
connectivity, which threatens to increase vulnerability to malicious
attacks. While details on network topology, protocols, and control differ between engineering domains, the fundamental architecture is similar. We now explain that architecture using an industrial control system (ICS).

A modern industrial control system typically consists of several
layers of networks. An example industrial control network is
illustrated in Fig.~\ref{fig:swatNetwork}. The physical process is
measured by distributed sensors, and manipulated by actuators. These
sensors and actuators operate by receiving and sending analog
signals. The analog signals are converted into digital signals by
Programmable Logic Controllers (PLCs). The digital signals are then
exchanged between PLCs and a central supervisory control system
(SCADA) using industrial communication protocols
(e.g. Modbus/TCP). 
  

\Paragraph{Modeling CPS} CPS can also be seen as a set of
communicating agents~\cite{cardenas09challenges} (often with
one node acting as a controller), and related work focuses on the
representation of the concrete behavior of the CPS~\cite{taormina16epanet,antonioli15minicps,adepu15attacker,urbina16fieldbus}. This
is believed to help the discovery the new attacks specific for CPS,
e.g. resonance attacks~\cite{cardenas09challenges}.  However, that can
lead researchers to over-complicating the system models even when
searching for security attacks.

\subsection{The Dolev-Yao Model}
The DY attacker model~\cite{dolev83attacker} is a de-facto standard
for the formal analysis of information security.  The usage of such an
attacker model is usually employed for the identification of
cyber-related attacks, e.g, Web applications and Service-Oriented
architectures as proposed in~\cite{armando12avantssar,csrf}.  Attacker
models \`a la DY have been
proposed~\cite{steinmetzer15lockpicking,Schaller09wireless} to reason
on CPS.
In this work, we consider the standard DY~\cite{dolev83attacker} model of an active attacker 
who controls the network but cannot break cryptography. 

The attacker 
can intercept messages and analyze them if he possesses the
corresponding keys for decryption, and he can generate messages from
his knowledge and send them under any agent name. As usual,
for a set $M$ of
messages, we define $\dy$ (for ``Dolev-Yao'' knowledge) to be the smallest set closed under the \emph{generation
\textit{(G)}} and \emph{analysis \textit{(A)} rules} of the
\emph{system} 
given in Fig.~\ref{fig:DY}. The $G$ rules
express that the attacker can compose messages from known messages
using pairing, asymmetric and symmetric encryption. The $A$ rules
describe how the attacker can decompose messages. 

\begin{figure*}[t]
  \begin{displaymath}\scriptsize
    \renewcommand{\arraystretch}{3}
    \begin{array}{c}
      \vcenter{\infer[\DYCompAxiom]{M_1 \in \dy}{M_1 \in
          M}} \qquad
      \vcenter{\infer[\DYCompPair]{\conc{M_1,M_2} \in
          \dy}{M_1 \in \dy & M_2 \in
          \dy}} \qquad
      \vcenter{\infer[\DYCompCrypt]{\acrypt{M_1}{M_2} \in
          \dy}{M_1 \in \dy& M_2 \in
          \dy}} \\
      \vcenter{\infer[\DYCompScrypt]{\scrypt{M_1}{M_2} \in
          \dy}{M_1 \in \dy & M_2 \in
          \dy}} \qquad
      \vcenter{\infer[\DYCi]{M_i \in \dy}{\conc{M_1,M_2}
          \in \dy}} 
      \qquad
      \vcenter{\infer[\DYSK]{M_1 \in \dy}{\scrypt{M_1}{M_2}
          \in \dy & M_2 \in \dy}} \\
      \vcenter{\infer[\DYPK]{M_1 \in \dy}{\acrypt{M_1}{M_2} \in
          \dy & \inv{M_2} \in \dy}} \qquad
      \vcenter{\infer[\DYPrK]{M_1 \in \dy}{\acrypt{M_1}{\inv{M_2}}
          \in \dy & M_2 \in \dy}}
    \end{array}
  \end{displaymath}
\caption{The system of rules of the Dolev-Yao attacker}\label{fig:DY}
\end{figure*}	

The \emph{algebra of messages}, which tells us how messages are constructed,
is defined following~\cite{BasinMoedersheimViganoIJIS05}, 
in the standard way. In this paper, we consider the following
operations:
\begin{compactitem}
\item $\acrypt{M_1}{M_2}$ represents the \emph{asymmetric encryption} of $M_1$ with public key $M_2$;
\item $\acrypt{M_1}{\inv{M_2}}$ represents the \emph{asymmetric encryption} of $M_1$ with private key $\inv{M_2}$ (the mapping $\inv{\cdot}$ is discussed below);
\item $\scrypt{M_1}{M_2}$ represents the symmetric encryption of $M_1$ with symmetric key $M_2$;
\item $\conc{M_1,M_2}$ represents the concatenation of $M_1$ and $M_2$. 
\item $\inv{M}$ gives the private key that corresponds to public key
$M$
\end{compactitem}

\section{The Dolev-Yao Model is not Enough}
\label{sec:motivation}
Although the classic DY model can be applied to CPS
security analysis straight away, we argue that it will not be able to
detect a large set of attacks possible in that context (i.e. those
that involve physical-layer interactions). To illustrate that
argument, we now provide three example scenarios.
For the sake of simplicity, we start by presenting the intuition
behind the model and the goal. Further details on the ASLan++
prototype of these scenarios along with our results are provided in
Section~\ref{sec:usecase} and~\cite{rocchetto16cpdytool}. 

\subsection{Application of Dolev-Yao for CPS}
\label{sec:application}
\begin{figure}[tb]
\centering
\begin{subfigure}[t]{0.35\linewidth}
\includegraphics[height=2cm]{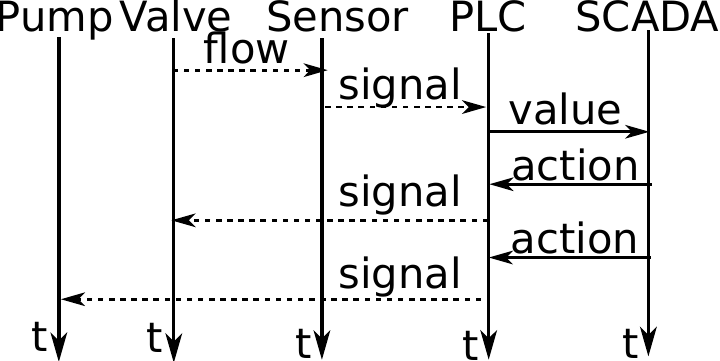}
\caption{}
 \label{fig:toyexample}
\end{subfigure}
\qquad
\begin{subfigure}[t]{0.35\linewidth}
\centering
\includegraphics[height=2cm]{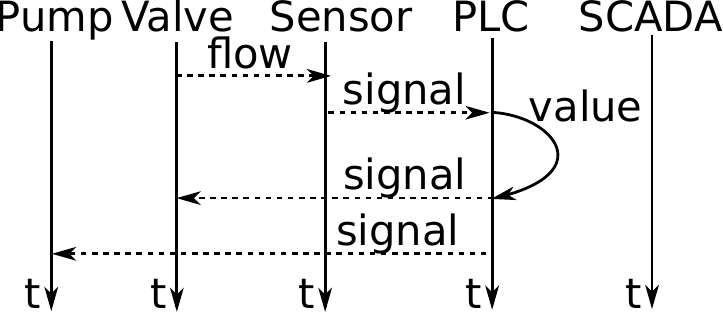}
\caption{}
 \label{fig:toyexample_limitations}
\end{subfigure}
\caption{(a) Physical interaction (dashed line) and digital interaction
  (solid line) between components in the first example. (b) Example interactions, with direct (out-of-band) communication between PLC, sensor, and actuator.}
\end{figure}

We base our example on a minimal setup in a water treatment system
(see Fig.~\ref{fig:process}). In particular, we use a subprocess of a
real water treatment testbed depicted in Fig.~\ref{fig:swatp1} (the SWaT
testbed~\cite{adepu15attacker,urbina16fieldbus}). A similar scenario has been
considered in~\cite{morris11control,reaves12open}.
The scenario we considered contains five different components and a PLC:
\begin{compactenum}
\item A \emph{motorized inflow valve}, initially open, let water flows into a tank through a pipe
\item A \emph{tank} is equipped with a \emph{sensor} which checks the
  level of the water inside the tank
\item The sensor communicates its reading of the level of the water
  inside the tank to a \emph{PLC}
\item When the level of the water reaches a certain upper threshold, the PLC
  communicates to the motorized inflow valve to close and to the \emph{pump} to start
\item Symmetrically, when the water reaches a certain lower threshold, the
PLC communicates to the inflow valve to open and to the pump to stop
\item A \emph{manual valve} (placed between the tank and the pump) can be manually
opened/closed, e.g., to prevent the water to flow into the rest of the testbed
in case the water in the tank is contaminated or the pump broken
\item A central \emph{SCADA} control that communicates with the PLC over the network
\end{compactenum}
In the following, we assume that the attacker's goal is to cause a
water spillage (or burst) in the tank component. The abstract messages
exchanged over the network are quite simple (see
Fig.~\ref{fig:toyexample}).  The valve controls the inflow of the
water to the tank.  The sensor of the tank reports the current fill
state to the PLC as analog signals.  The PLC converts the analog
signals into digital messages (value in the picture) that it sends to
the SCADA.  If the water level in the tank has crossed certain
high/low thresholds, the SCADA sends a close/open message to the
inflow valve and on/off to the pump.  We note that in this setting,
there is no distinction between the tank and the sensor.  Nothing
prevents us in considering them as two separate entities, but this
would complicate the model without benefit in terms of attacks.

We analyzed the ASLan++ model using the AVANTSSAR platform, and found
a simple attack (as expected). In that attack, the attacker drops the
messages from the 
PLC to the SCADA. As result, the tank will
overflow and the attacker will achieve his goal. Even if the attack
is simple, it shows that one can easily use state-of-the-art
verification tools and the standard DY attacker model to search for
attacks on a CPS model.
This basic example demonstrates that it is possible to reason on
similar scenarios without considering the (fluid)
mechanics of the process under attack. In particular, even in such a
simple example the equations describing the flow of the water are far
from trivial and are usually considered when reasoning on similar
scenarios. In the remainder of this section, we consider a
modification of this scenario in which the network-only DY attacker model cannot
find the attack.

\subsection{Limitations of Dolev-Yao for CPS}

The previous examples demonstrates that the DY attacker model can be
used to reason on network-related security aspects of CPS. We now
consider a scenario in which the agents also interact through
physical-layer interactions. In particular, in real-world CPS the
communication between PLCs, sensors and actuators usually uses analog
signals. When a distributed control scheme is used, the logic of the
system is usually integrated directly into the PLC, instead of relying on
the SCADA component.  In addition, components can interact though the
physical layer directly (e.g. by exchanging water from a pipe into a
tank).
To reflect such a setting, we modify the system in the previous
example as follows: the inflow valve and the pump are now directly controlled by the PLC
based on analog signals from the sensor. In other words, the valve and the pump
operate without interrogating the SCADA.  As a result, the messages
on the network and SCADA are not directly involved in the operations
of the valve and the pump. 
We can consider the setting as one in which there is
just one entity whose internal behavior encapsulate the behaviors of
the valves, sensor, tank, pump and PLC (see
Fig.~\ref{fig:toyexample_limitations}). Since there is no
communication over the network related to the opening/closing of the
valve or to the level of the water, there is no way for the DY
attacker model to achieve the goal, i.e., overflowing the
tank. We confirmed this intuition with a related
model in ASLan++ and evaluation in the AVANTSSAR platform. No
successful attack is found.


Nevertheless, it could be expected that attacks by a physically
present attacker are possible in the given setting, in particular if a
physically present attacker can manually open or close the valves. Our
ASLan++ model does not find such an attack because potential malicious
physical-layer interactions with the system have not been considered.


\subsection{Proposed Approach}
We claim that, (so far) the related work 
generally models the operation of a CPS as a set of messages exchanged
between entities over a network (see
Section~\ref{sec:related}). For that reason, we consider prior work as
limited modification of the DY attacker model. However, consideration
of the physical actions is often crucial to find real-world attacks on
CPS (e.g. attacks such as Stuxnet~\cite{weinbergerStuxnet}). For that
reason, we propose an extension of the DY attacker model with new
physical interaction rules to support reasoning on the physical-layer
security of CPS. We will introduce those rules in
Section~\ref{sec:extension}. Before that, we briefly discuss two
aspects of our proposed approach in more detail: abstraction of physical process behavior, and whether verification tools or model checkers are better suited.

\Paragraph{Abstraction of Physical Processes} 
In our proposed approach, physical layer interactions will be modeled
as abstract interactions between components. In particular, we do not
model all the details of the behavior of an agent for CPS. We believe
that it will be very challenging for a security verification tool (or
a model checker in general) to consider all details of the behavior of
an agent for CPS.  For example, differential equations that model the
behavior of an ultra-filtration process will be difficult to consider
by the DY model or verification tool. 

In this work, we abstract away all these details, similar to the way
that perfect cryptography is used for security protocols analysis
where we abstract from cryptographic primitives
(see~\cite{cortier2006survey} for more details).  In security
protocols, that abstraction is justified by the observation that most of the
attacks rely on the logical aspects of the protocol. Encryption
schemes are treated as black box and the attacker cannot learn any
useful information from an encrypted message without the proper
decryption key. As such, a generic predicate over a term defines the
encryption as $\acrypt{M_1}{M_2}$ in Section~\ref{sec:background}. In
CPS, we assume that all the physical processes can be abstractly
represented.  

\Paragraph{Verification Tools vs. Model Checkers}  In particular, we propose to use a DY verification tool, and
not a general model checker.  Our argument for that is the following:
in order to model a CPS to formally validate it against an attacker
model, CPS (or subparts) are often modeled with languages supported by
tools which do not implement the DY attacker model, e.g. NuSMV,
SPIN~\cite{Choi07nusmv}.  However, it has been
shown~\cite{basin2011model} 
that an ad-hoc implementation of the DY is more advanced in terms of
efficiency and coverage than using a ``general purpose'' model checker
with the DY model-hard coded in the specification.
That is particularly evident when considering the numerous
amount of verification tools developed specifically to reason on 
the security aspects of various systems, e.g., \cite{blanchet2001proverif,BasinMoedersheimViganoIJIS05,ArmandoCompagna-jelia04-sd,turuani06atse,escobar2009maude}.

\section{Physical-Layer Interactions for the Dolev-Yao Attacker}\label{sec:extension}

In this section, we present our proposed extensions of the DY 
model in order to make it suitable to argue about security of
Cyber-Physical Systems. In particular, we discuss the introduction of additional rules for the DY attacker model to describe physical-layer interactions.

\subsection{New Rules for the DY Attacker and System}
\label{sec:rules}

\Paragraph{New Rules for the System} The new rules for our system
model aim to capture the diverse physical-layer interactions between
components in the system under attack. The interactions are usually
constrained by the laws of physics, which will never be violated. A
very exhaustive coverage of all kinds of physical layer interactions
and laws of physics would potentially result into a large set of
additional rules (potentially automatically extracted from a system
specification, e.g.,~\cite{schmidt09distilling}). In the following, we will consider only few additional
rules to model specific interactions. In Fig.~\ref{fig:physicsSystem}, we
present rules that represent laws of physics related to our
example (Fig.~\ref{fig:process}).  With a slightly abuse of
notation, each rule represents a modification of the system status
$\sysstat$ from preconditions (top) to postconditions (bottom).
$\sysstat$ is a set collecting all the physical properties of the
systems (e.g., water level, temperature, pressure) for each component in the system
(e.g., tank). The properties are expressed with the predicate
$\mathit{C(property,value)}$ (C($\cdot$) as short form of Component($\cdot$)).
In Fig.~\ref{fig:physicsSystem},
$\mathit{raise_1(Tank)}$ and $\mathit{raise_2(Tank)}$ 
relates a system configuration with its physical effects,
i.e.,
the increase of the water level in the tank.
$\damage$ expresses the effect of the burst of the tank, and
$\closecomp$ and $\opencomp$ defines the effect of physical interactions
with a component (e.g., a valve) which can be manually operated to change its status.

\begin{figure*}[t]
\scriptsize
  \begin{displaymath}
    \renewcommand{\arraystretch}{3}
    \begin{array}{c}
      \vcenter{\infer[\mathit{raise_1(Tank)}]
      {\mathit{Tank(level,value')}\in \sysstat \wedge (\mathit{value'}>\mathit{value})}
      {\mathit{Tank(level,value)}\in \sysstat & \mathit{Pump(status,off)} & \mathit{InflwoValve(status,open)} \in \sysstat}}\\

      \vcenter{\infer[\mathit{raise_2(Tank)}]
      {\mathit{Tank(level,value')}\in \sysstat \wedge (\mathit{value'}>\mathit{value})}
      {\mathit{Tank(level,value)}\in \sysstat & \mathit{ManualValve(status,close)} \in \sysstat & \mathit{InflowValve(status,open)} \in \sysstat}}\\

      \vcenter{\infer[\damage]
      {\mathit{C(level,value')}\in \sysstat \wedge (\mathit{value'}<\mathit{value})}
      {\mathit{C(status,damaged)}\in \sysstat & 
       \mathit{C(contains,water)}\in \sysstat & 
       \mathit{C(level,value)}\in \sysstat}} \\

      \vcenter{\infer[\closecomp]
      {\mathit{C(status,close)}\in \sysstat}
      {\mathit{C(operate,manual)}\in \sysstat & \mathit{C(status,open)}\vee\mathit{C(status,close)} \in \sysstat}}\\

      \vcenter{\infer[\opencomp]
      {\mathit{C(status,open)}\in \sysstat}
      {\mathit{C(operate,manual)}\in \sysstat & \mathit{C(status,open)}\vee\mathit{C(status,close)} \in \sysstat}}\\
    \end{array}
  \end{displaymath}
\caption{Examples of rules that represent physical-layer interactions in the system}\label{fig:physicsSystem}
\end{figure*}

\Paragraph{Rules for DY Attacker} The new rules for our attacker model
aim to capture the diverse physical-layer interactions between the
attacker and the system (see
Fig.~\ref{fig:physicsAttacker}). Similar to the system specification
rules, the interactions between attacker and system are usually
constrained by the laws of physics---even a strong attacker would not
be able to create or consume arbitrary amounts of energy, move at
infinite speed, or similar.



\begin{figure*}[t]
\scriptsize
  \begin{displaymath}
    \renewcommand{\arraystretch}{3}
    \begin{array}{c}
      \vcenter{\infer[\mathit{damage}_\dya]
      {\mathit{C(status,damaged)}\in \sysstat}
      {\mathit{DYProp(distance,physical\_access)} & \mathit{DYProp(tool,damage)}}}\\

      \vcenter{\infer[\mathit{manualClose}_\dya]
      {\mathit{C(status,close)}\in \sysstat}
      {\mathit{DYProp(distance,physical\_access)} & 
		\mathit{C(operate,manual)}\in \sysstat & \mathit{C(status,open)} \in \sysstat}}\\

      \vcenter{\infer[\mathit{manualOpen}_\dya]
      {\mathit{C(status,open)}\in \sysstat}
      {\mathit{DYProp(distance,physical\_access)} & 
		\mathit{C(operate,manual)}\in \sysstat & \mathit{C(status,close)} \in \sysstat}}\\
     \end{array}
  \end{displaymath}
\caption{Examples of rules that represent physical-layer capabilities of attacker}\label{fig:physicsAttacker}
\end{figure*}	

\subsection{Implementation of New Rules for DY}
We base our attacker model on a review of related work that
aims to profile attackers for
CPS~\cite{cardenas09challenges,Cardenas09rethinking,urbina16fieldbus,leMay11advise}.
We found that they all share the idea
of defining the attacker by means of a set of \emph{dimensions}. These
dimensions can be seen as \emph{properties of the attacker}, e.g.,
distance with respect to the CPS, knowledge of the physics of the
components of the system, tools (software and hardware) available to
the attacker, financial support, and preference to stay hidden.  We
can use dimensions together with \emph{physical properties of the
  system} to define new rules for the attacker as follows.

\begin{displaymath}
\begin{array}{c}
      \vcenter{\infer[\mathit{action}]{\mathit{result~of~action}}
		{\mathit{attacker\_property} & \mathit{system\_property}}} 
\end{array}
\end{displaymath}
where one or more attacker's properties along with the knowledge
of one or more system property (that might be related to the knowledge of 
some physical laws connected to the system property) 
are the precondition to perform an 
action \emph{action} which results are expressed as postconditions.
In other words, an attacker is a malicious agent that can take advantage
from the improper use of some device of the system. 

In our first example, we can add the $\mathit{damage}_{\dy}$ in
Fig.~\ref{fig:physicsAttacker} which express that an attacker who has
physical access to the CPS could damage or manually operate a
component, for example, a tank.  Other examples are rules expressing
that if the attacker has physical access to the CPS can (as we will
discuss in Section~\ref{sec:heating}) heat the tank and increase its
pressure. In Section~\ref{sec:usecase}, we will show how we can
leverage those attacker rules to find new attacks on a CPS which
involve physical-layer interactions.

\subsection{DY Rule Extension Using Horn Clauses}

In order to apply our idea to a concrete example, we require a
verification tool such that: (i) allows modification to the DY rules,
and (ii) provides a language expressive enough to model a CPS. It is
not easy to find a security verification tool with such constraints
and, to the best of our knowledge, there is no tools in the
literature. 

In this work, we propose the following two workarounds that allow us to
implement our additional rules even without a tool that satisfies the
mentioned requirements: (i) we have used Horn Clauses (HC) to add extra rules to the DY attacker model, and (ii) we have used databases (shared memories) to store the state of the components the system, e.g., the level of the water of a tank. Using both ideas, we require only a tool that supports Horn
Clauses and Shared Memories. We chose the ASLan++ specification
language~\cite{ASLan++-FMCO10} which supports both
HC and shared memories. Using ASLan++, we have implemented
several case studies and obtained preliminary results that
support our proposed approach.


\section{Case Studies}
\label{sec:usecase}

\label{sec:hc}
\begin{figure}[t]
\centering
\includegraphics[width=0.6\linewidth]{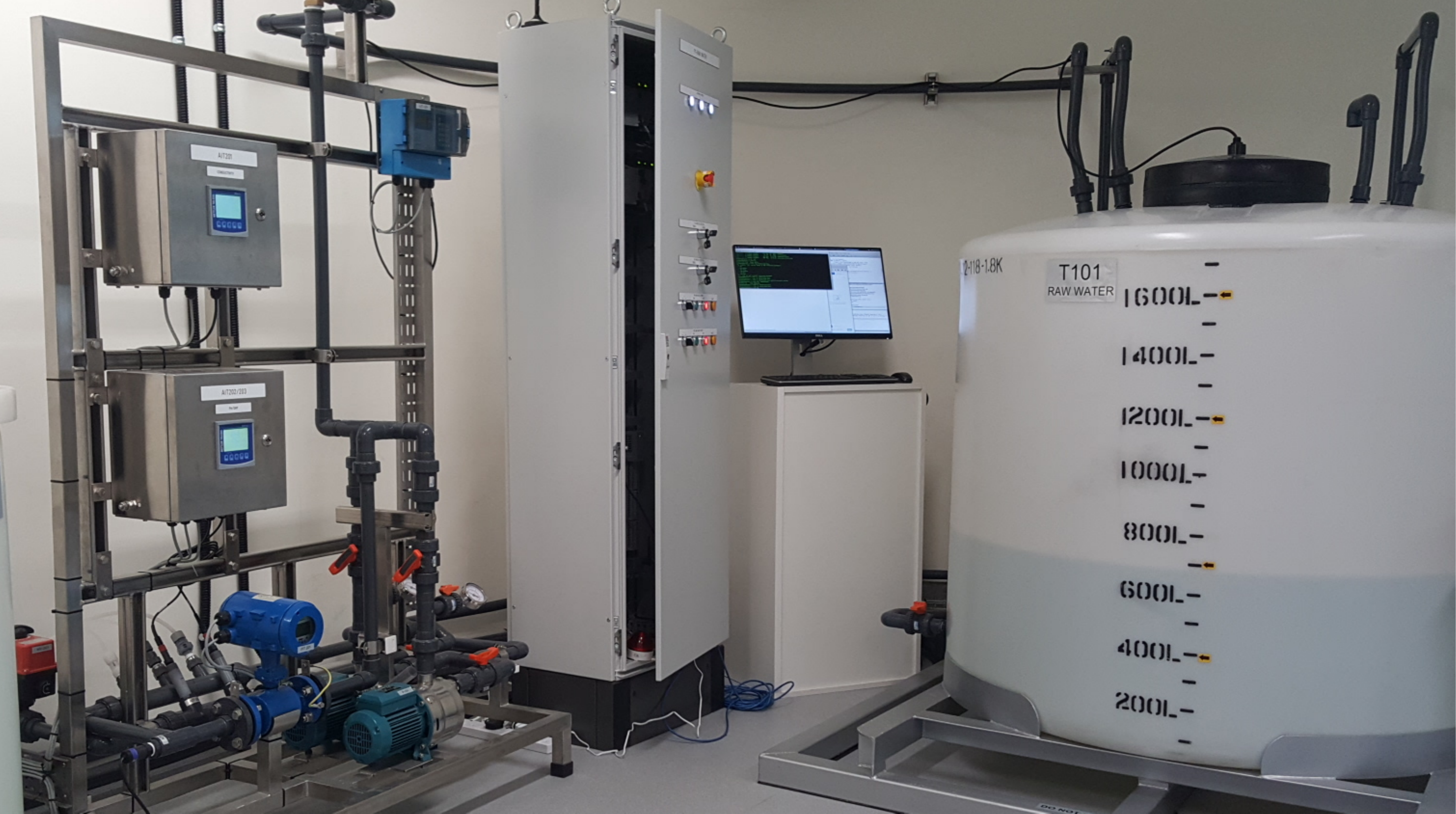}
\caption{Process P1 (raw water treatment) -- SWaT testbed}
 \label{fig:swatp1}
\end{figure}

In this section, we first show that a DY verification tool can be used
to check security goals in CPS models. We have used the ASLan++
specification language~\cite{ASLan++-FMCO10} to define our
examples 
which are based on a process of the SWaT
testbed~\cite{mathur16swat} depicted in Fig.~\ref{fig:swatp1} (see
Fig.~\ref{fig:toyexample} for the message sequence chart). We
start by providing more details on the example summarized in
Section~\ref{sec:motivation}. In particular, we present a network-only
modeling of the CPS and potential attacks, which shows that the DY
model can be used to find attacks similar to ones discussed in related
work (e.g., man-in-the-middle attacks in
\cite{antonioli15minicps,urbina16fieldbus}).  For that analysis, we
abstract away the implementation details of the CPS and detect the
same (network related) security flaw of most of the approaches we have
found in the literature. We then modify the specification (as depicted in Fig.
\ref{fig:toyexample_limitations}) to show that when some physical
operations (which are the at very core of a CPS) are involved in the
process, the standard DY attacker model might not 
be able to find all attacks. 
To mitigate this, we modify the DY model to let him physically interact
with the system under certain constraints.  This allows the attacker
to detect new attacks which involves physical interaction with the
system.  We show that almost all the attacks which relay on attackers'
physical actions 
cannot be found. We propose a first investigation
on how to extend the DY model in order to capture both cyber and
physical attacks.
Our results, along with timing, are summarized in Table~\ref{table:results}.
\begin{table}[t]
\centering
\caption{Summary of the analysis on the use cases\label{table:results}}
\begin{tabular}{lccll}
&
\multicolumn{2}{c}{attack found} &
\multicolumn{2}{c}{timing}\\
\toprule
&
\multicolumn{1}{c}{DY}&
\multicolumn{1}{c}{CPDY}&
\multicolumn{1}{c}{analysis}&
\multicolumn{1}{c}{total}
\\
\midrule
Network (Section~\ref{sec:swat}) &$\checkmark$&$\checkmark$& 220ms &1.7s \\
Manual (Section~\ref{sec:swat-physical}) &&$\checkmark$& 8ms &1.3s\\
Heating (Section~\ref{sec:heating}) &&$\checkmark$& 4ms & 1.0s\\
\bottomrule
\end{tabular}

{\centering total includes time for translation, analysis and attack trace generation}
\end{table}

\subsection{Network-Based Communication Use Case}
\label{sec:swat}
This section briefly summarizes an implementation of the scenario
proposed in Section~\ref{sec:application}. We focus on three aspects:
\begin{compactitem}
\item The \emph{status} of the system (e.g., the level of water and measurements of sensors)
\item The \emph{behavior} of each entity (i.e., tank, valves, pump, PLC and SCADA)
\item The \emph{communication} between various entities (analog and network channels)
\end{compactitem}
In this example, the PLC converts the analog signals to digital
messages and sends them to the SCADA control. To be coherent with the
example, we model an analog channel (e.g., by using a database)
between the inflow valve, the tank, the pump, and the PLC. The PLC then translates and
communicates the tank/valve/pump status over a network channel with the
SCADA.  For the sake of readability, we assume the PLC automatically
converts and sends the tank/valve/pump status. As a result, the valve,
the tank, and the pump directly communicate with the SCADA over a network
channel. The full implementation is reported in~\cite{rocchetto16cpdytool}. 

\paragraph{System Status.}
CPS can be seen as communicating over two channels: one is the network
channel (maybe itself divided into several layers or regions) and the
other is the physical flow of the events, e.g., electricity in power
grids or water in water treatment or distribution CPS.  We believe
that an understanding on how to model the physical flow of a CPS,
integration of that with the network (in such a way that an attacker
model can concretely find new attacks) is still not well defined in
the literature.

In this work, the status of the system is defined 
by a database $\mathit{systemStatus}$, shared between all the entities
(but hidden to the DY attacker). The database is defined as a set of pairs
$\mathit{(agent,status)}$ that keeps track of the status of all entities of the specification.



\paragraph{Inflow Valve/Pump.}
The valve/pump specification describes a scenario where a SCADA changes the
status of the valve/pump, e.g., from open/on to close/off, by sending to the valve
a message through the PLC.

\paragraph{Manual Valve.} 
The behavior of the manual valve is the same as the inflow valve. The only
difference is that the manual valve can only be manually operated (e.g, to change its status from open to close), i.e., cannot be operated using network messages.

%
%


\paragraph{Tank.}
In the real testbed the PLC 
interrogates the sensor of the tank in order to
obtain the level of the water inside the tank. For simplicity, 
we do not distinguish between the tank as a container and its sensors.
We also assume that the sensor sends the sensed data of the level of the water
whenever the level is above or below a certain threshold. 
We can obviously consider the more complicated and realistic 
tank specification 
containing a sensor
that waits for the PLC 
to interrogate it. This complicates the analysis 
but the performance of the validation phase does not change order of magnitude 
(there is a variation of some milliseconds) and the result of the analysis remains
the same. We recall that messages are directly sent to the SCADA
instead of PLC for readability.

The tank model checks for two, mutually exclusive, status of the tank.

\begin{compactitem}
\item If the level of the water has reached an upper threshold $\mathit{overT}$,
the inflow valve is closed, the pump is on, 
and the manual valve is open,
the $\mathit{systemStatus}$ database is updated
as if the water level had lowered to a lower threshold $\mathit{underT}$.
After the status update, 
the tank 
communicates its new status $\mathit{underT}$ to the SCADA.

\item Symmetrically, if the level of the water has reached a threshold $\mathit{underT}$,
the inflow valve is open, 
and either the manual valve is closed or the pump is off,
the $\mathit{systemStatus}$ database is updated
as if the water level had reached the upper threshold $\mathit{overT}$.

\end{compactitem}

%

\paragraph{SCADA.}
As for the valve entity, 
we have defined the behavior of the SCADA
waiting for incoming messages from the tank entity. 
When the tank communicates to the SCADA
that the water has reached the upper 
threshold $\mathit{overT}$,
the SCADA
closes the inflow valve and turns on the pump. 
Symmetrically, when
the tank reaches the lower threshold, 
the SCADA
opens the inflow valve and turns off the pump. 


\paragraph{Initial Status of the System.}
The initial status of the specification 
is defined 
with the tank empty 
(i.e.,
the level of the water is $\mathit{underT}$ in the $\mathit{systemStatus}$ database),
the inflow and the manual valve are open, and the pump is off. 
%

\paragraph{Goal.}
The goal is to overflow the tank and in ASLan++ we can define our goal
as the following LTL (Linear Temporal Logic) formula. 
\footnotesize
\begin{gather*}
\Box (\mathit{inflowValve}(\mathit{status},\mathit{open})\in\mathit{Sys} \Rightarrow\\
 \mathit{manualValve}(\mathit{status},\mathit{open})\in\mathit{Sys} \wedge 
 (\mathit{tank}(\mathit{status},\mathit{underT}) \in \mathit{Sys} \vee \mathit{pump}(\mathit{status},\mathit{on}) \in \mathit{Sys})) 
\end{gather*}
\normalsize
In the formula, we define that whenever the inflow valve is open, i.e. the $\mathit{systemStatus}$
database contains $\mathit{valve}(\mathit{status},\mathit{open})$, 
then the manual valve is open, and
either the tank must be empty or the pump turned on. 
In other words, if we find a configuration of the system such that the
inflow valve is opened, the tank is full of water, and the pump is off, then the tank is overflowing.
The $\Box$ at the beginning of the goal states that the goal must
hold in every state of the system (i.e., LTL global operator).



\paragraph{Security Analysis.}
The AVANTSSAR platform finds a violation of the goal (i.e., a states
where the goal does not hold). 
The goal is violated because there is a
state of the system in which the tank has reached the
$\mathit{overT}$ but the valve is still open and the pump is switched off.
In order
to achieve the goal, an attacker have to drop the packet,
communicating the $\mathit{overT}$ status, sent from the tank to the
SCADA. 


\subsection{Physics-Based Interaction Use Case}
\label{sec:swat-physical}
We now modify the scenario by removing the communication of the
level of the water between the PLC and the SCADA, i.e., between
tank/valve/pump and SCADA in the previous specification.  For that reason,
we assume that the PLC automatically close the inflow valve when the level of
the water inside the tank reaches the threshold level
$\mathit{overT}$.  The DY attacker cannot spoof or eavesdrop
  the communication between entities anymore since there is no more
  network communication with the SCADA.

\paragraph{Security Analysis.}
Against the DY model, the AVANTSSAR platform does not
report any attack on the specification with respect to the 
goal defined in Section~\ref{sec:swat}.
This result is straightforward since the attacker does 
not receives any message and there is no interactions over network
between various entities.

As we are considering a CPS, an attacker who could have physical
access to the system could most likely find a number of ways to
overflow the tank. Being in close proximity of the CPS
could give to the attacker an advantage with respect to a
cyber-attacker who can only access the system through the network.
For example, an attacker could manually operates the valves 
to increase the level of the water in the tank and burst the tank.
In this perspective, it is fair to assume that
there are some attacker properties, e.g., distance, that can be
exploited by an attacker to perform some actions (or even attacks)
that leads to some physical consequences in the CPS.
As a simple motivating example, the following rule 
has been added to the specification (as a Horn clause as discussed in Section~\ref{sec:hc}).
\begin{displaymath}\footnotesize
\begin{array}{c}
      \vcenter{\infer[]
	{\mathit{C}(\mathit{status},\mathit{close})\in \mathit{Sys}}
	{\mathit{DYProp}(\mathit{distance},\mathit{physical\_access}) \wedge \mathit{C}(\mathit{status},\mathit{open})\in \mathit{Sys} \wedge \mathit{C}(\mathit{operate},\mathit{manual})\in \mathit{Sys}}
} 
\end{array}
\end{displaymath}
The clause states that whenever the attacker has physical access to the
CPS, 
he can close any valve which can be manually operated.
We also modeled the opening of a valve.

	


When we run the AVANTSSAR platform searching for a state of the system in 
which the level of the water in the tank has reached the upper threshold $\mathit{overT}$ (defined as a goal), 
we (unsurprisingly) find an attack.
To perform the attack, the attacker manually opens the inflow valve and closes the manual valve.

\subsection{Physics-Based Interaction Use Case -- a Stronger Attacker}
\label{sec:heating}
This use case is a variation of 
the one  in Section~\ref{sec:swat-physical}, where we add extra rules to the attacker.
In particular, we have modeled that whenever
the attacker has physical access to the CPS, he can
physically interact with the system
and heat up the water in tank (e.g., through microwave or fire).
\begin{displaymath}\footnotesize
\begin{array}{c}
      \vcenter{\infer[\mathit{heat}_\dya]
	{\mathit{C(status,heating)}\in \sysstat}
	{\mathit{DYProp(Distance,physical\_access)} & \mathit{DYProp(Tool,heating)}}
} 
\end{array}
\end{displaymath}
Note here that some system properties should hold, e.g., 
the tank should contain a liquid but for the sake of simplicity we abstract
away system properties in this example.



We have modeled the corresponding physical laws and we report three
examples in Fig.~\ref{fig:physics_heat}.  $\tinc$ and $\pinc$ express
the temperature and pressure increase when heating a generic component
containing water respectively.  $\ptinc$ defines the direct
proportionality between temperature and pressure in the presence of
water inside a component.
 \begin{figure*}[t]
 \scriptsize
   \begin{displaymath}
     \renewcommand{\arraystretch}{3}
     \begin{array}{c}
       \vcenter{\infer[\tinc]
       {\mathit{C(temperature,Level')}\in \sysstat \wedge \mathit{Level'}>\mathit{Level}}
       {\mathit{C(status,heating)}\in \sysstat & \mathit{C(contains,water)}\in \sysstat & \mathit{C(temperature,Level)}\in \sysstat}} \\

       \vcenter{\infer[\pinc]
       {\mathit{C(pressure,Level')}\in \sysstat \wedge \mathit{Level'}>\mathit{Level}}
       {\mathit{C(status,heating)}\in \sysstat &\mathit{C(contains,water)}\in \sysstat &\mathit{C(pressure,Level)}\in \sysstat }} \\

       \vcenter{\infer[\ptinc]
       {\mathit{C(temperature,TLevel')}\in \sysstat \wedge \mathit{C(pressure,PLevel')}\in \sysstat \wedge (\mathit{PLevel'}>\mathit{PLevel}) \wedge (\mathit{TLevel'}>\mathit{TLevel})}
       {\mathit{C(status,heating)}\in \sysstat &\mathit{C(contains,water)}\in \sysstat & \mathit{C(temperature,TLevel)}\in \sysstat & \mathit{C(pressure,PLevel)}\in \sysstat}} \\

     \end{array}
   \end{displaymath}
 \caption{Rules that represent physical laws in use case of Section~\ref{sec:heating}}\label{fig:physics_heat}
 \end{figure*}	

\paragraph{Goal.}
We check if the attacker can burst the tank, increasing the pressure of the tank.
\footnotesize
\begin{displaymath}
\Box (\mathit{Tank}(\mathit{pressure},\mathit{overT})\not\in\mathit{Sys}) 
\end{displaymath}
\normalsize



\paragraph{Security Analysis.}

The AVANTSSAR platform reports a violation of the 
goal. 
The two clauses $\mathit{heat}_2$ and $heat_{\dy}$
have been used 
to heat the tank 
component and then to raise its pressure, bursting the tank. 

\section{Related Work}
\label{sec:related}

The formal verification of security properties of CPS is a non trivial task, as 
CPS introduce physical properties to the 
system under analysis. 
SAT/SMT solvers used by security analysis tools 
(e.g.,\cite{armando12avantssar}) do not support such properties.
In order to overcome this limitation, one could simulate
 the process (e.g.,~\cite{antonioli15minicps}) or adapt
the level of abstraction of CPS components.
In~\cite{vigo12attacker}, the author presents a formal definition 
of an attacker model for CPS.
%
The attacker is defined as a set of pairs representing locations and
capabilities. Capabilities are defined as a set of tuples expressing
actions, cost (energy/time) and range (with respect to the topology)
of the attacker. The attacker is assumed to perform two types of
attacks: \emph{physical}, against a device and \emph{cyber} against
the communications; where the first requires physical access while the
second proximity to the node.  The actions of the attacker
are \emph{WSN actions} (remove, read/write, reveal, reprogram, starve) and
\emph{cyber actions} (block, eavesdrop, inject).


In~\cite{Schaller09wireless,basin11formal}, the authors present a
formalization to reason on security properties of wireless networks
(including a considerations of physical properties related to those
networks). The authors present an attacker model as a variation of the DY
attacker model. The attacker is a malicious agent of the network who
cannot break cryptography. 
He has a
fixed location, while the usual DY controls the entire network, a set
of transmitters and receivers, an initial knowledge with his
private/public keys which can use to create and analyze messages.
The authors also consider constraints on the distance of communicating
parties. An attacker can only intercept messages at his location and
colluding attackers do not instantaneously exchange knowledge, they
are constrained by the network topology.

\section{Conclusions and Future Work}
\label{sec:conclusion}
In this paper, we argued that (to the best of our knowledge) current
approaches for the formal reasoning on the security of CPS do not
consider most of the physical interaction between the attacker and the
system.  Instead, the works we reviewed only focus on the network
interaction between components of a CPS, which is indeed important but
not sufficient for an extensive security analysis.  One of the main
difficulties of considering physical interaction of the CPS is that
usually this leads to the definition of the physical processes of
various components of the CPS. 

We proposed several basic uses cases in which the physical behavior of
both the CPS and the attacker can be used to produce attacks that rely
on physical actions of the attacker that are outside the normal
behavior of the CPS.  To alleviate that problem, we presented the idea
of extending the DY attacker to a CPDY attacker model that allows to
include physical-layer interaction.  We implemented that CPDY model in
ASlan++ and used the AVANTSSAR platform to show that our extended
attacker model is indeed able to discover the physical-layer attacks
in CPS.



\Paragraph{Acknowledgments} This work was supported by the National Research Foundation of Singapore under grant NRF2014NCR-NCR001-40.

\bibliographystyle{abbrv}
\bibliography{../../bibs/bibliography}
\end{document}